\documentclass[prl,twocolumn]{revtex4}
\usepackage{epsfig}
\usepackage{bm}
\usepackage{amsmath}

\begin{document}

\title{Predictability problem in dynamical systems and in chaotic climate dynamics}

\author{A. Bershadskii}

\affiliation{
ICAR, P.O. Box 31155, Jerusalem 91000, Israel
}

\begin{abstract}
  Predictability horizon properties of chaotic dynamical systems can be related to their spectral properties. It is shown, using this relationship, that the spectral properties of the leading large-scale climate daily indices indicate a smooth predictability: i.e. their intrinsic predictability horizons can be indefinitely extended by reducing the initial error. Special properties of the Madden-Julian Oscillation and some models (including ensemble weather forecasting) have been also discussed in this context.

\end{abstract}

\maketitle

\section{Power spectrum and predictability}

  A simple and assured way to distinguish between chaotic and quasi-periodic dynamics is to study power spectrum of the time signals generated by the dynamical systems. For chaotic systems the power spectrum will be a {\it broadband} one. Figure 1, for instance shows a power spectrum computed for $z$-component of the Lorenz equations  \cite{loren}. 
$$
\frac{dx}{dt} = \sigma (y - x),~~      
\frac{dy}{dt} = r x - y - x z, ~~
\frac{dz}{dt} = x y - b z      \eqno{(1)}          
$$
for the parameters The $b = 8/3, ~\sigma=10.0$ and $r = 28.0$. The power spectrum was computed by the maximum entropy method (with an optimal resolution \cite{oh}). The {\it broadband} spectrum decays exponentially with frequency (dashed straight line in the semi-log scales)
$$
E(f) \propto \exp -(f/f_c)      \eqno{(2)}
$$
 
\begin{figure} \vspace{-0.6cm}\centering
\epsfig{width=.45\textwidth,file=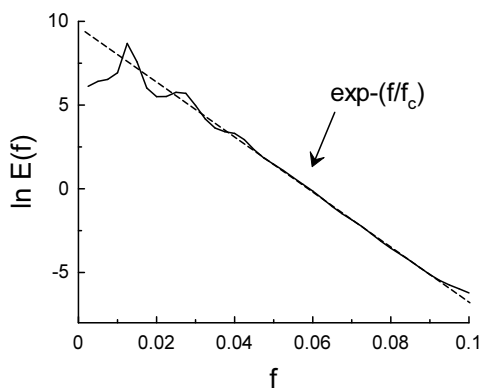} \vspace{-4.5cm}
\caption{Logarithm of power spectrum vs. frequency $f$. The dashed straight line indicates Eq. (2).} 
\end{figure}
  It is also true for many other dynamical systems with the chaotic behaviour (including some atmospheric phenomena) \cite{oh}-\cite{oa}. This spectral decay can be related to the {\it smooth} sensitivity  to the initial conditions mentioned by E. N. Lorenz in his 1969 paper \cite{l2}: 
  
  "The error eventually becomes much larger than the initial error. At any particular future time the error may be made arbitrarily small by making the initial error sufficiently small, but, no matter how small the initial error (if not zero), the error becomes large in the sufficiently distant future."
\begin{figure} \vspace{-0.5cm}\centering
\epsfig{width=.45\textwidth,file=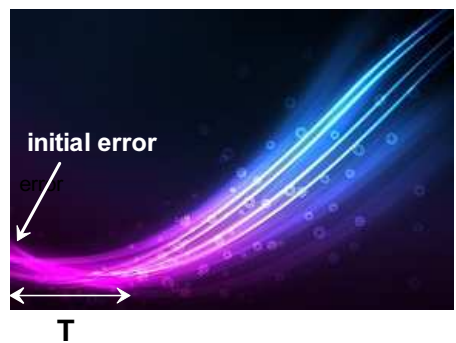} \vspace{-5.3cm}
\caption{A sketch of the trajectories development.} 
\end{figure}
  
  The initial error (say an $\varepsilon$-ball in the phase space, with the initial state of the trajectory as its center) is a measure of the differences in the initial conditions. The above mentioned smoothness of the chaotic behaviour is naturally related to the predictability problem (predictability horizon properties). Figure 2 shows a sketch of development of the trajectories of a chaotic system. For the smooth case one can make the time of controlled behaviour {\bf T} arbitrary large (but still finite) by reducing the initial error. The {\bf T} can be also considered as a predictability horizon.
   
   Some trajectories starting from the error's $\varepsilon$-ball can violate the controlled behaviour, but measure corresponding to these trajectories will be zero.\\

   Let us define (following the Ref. \cite{l2}) a {\it smooth} and a {\it rough} predictability. At smooth predictability  the predictability horizon can be indefinitely extended by reducing the initial error. 
 
   At {\it rough} predictability the predictability horizon cannot be indefinitely extended by reducing the initial error.\\
   
   The Lorenz system Eq. (1), for instance, has the smooth predictability and the exponential spectral decay Fig. 1. This relationship can be generalized for a wide class of the chaotic systems. Namely, the chaotic systems with the smooth sensitivity to the initial conditions (the smooth predictability) have a stretched exponential spectral decay:
$$
E(f ) \propto \exp-(f/f_0)^{\beta}  \eqno{(3)}
$$
where  $\beta \leq 1$. Moreover, for the chaotic systems with the rough predictability the spectral decay has a power-law form
$$
E(f) \propto f^{-\gamma}    \eqno{(4)}
$$
This dependence of the spectral decay on the smoothness resembles relation between smoothness and truncation of the Fourier transform to finite interval (do not confuse the smoothnes of a trjectory itself and the above described Lorenz's smoothness of the collective trajectories evolution in the phase space). In this resemblance the Lorenz's smoothness should be also extended on derivatives of all orders.\\

   For Hamiltonian systems with the smooth (in the Lorenz sense) behaviour only two values of the parameter $\beta$ are possible $\beta =3/4$ and $\beta = 1/2$ - distributed chaos \cite{b3} (but it should be noted that not all chaotic Hamiltonian systems have the spectral decay in the form of the Eq. (3) or Eq. (4)).\\

  Appearance of the stretched exponential spectrum Eq. (3) is a good indicator of the smooth predictability, whereas the power-law spectrum Eq. (4) indicates the rough predictability  (cf. the recent reviews Refs. \cite{ghil}-\cite{li}, next Section and the Appendices).

\section{Hamiltonian dynamical systems}

   Let us consider two examples of the Hamiltonian dynamical systems - one with smooth and another with rough chaotic behaviour.\\ 
  
  {\bf A}. Classic Toda lattice. \\
  
  The classic Toda lattice is a one dimensional $N$-particle Hamiltonian dynamical system having an exponentially decaying (with the distance between the particles) potential \cite{toda}. This dynamical system is completely integrable. 
  
  Figure 3 shows spectral data taken from Ref. \cite{els}, where the direct numerical simulations were performed for the classic Toda lattice on a ring with zero total momentum, periodic boundary conditions and random initial conditions. The appropriate scales were chosen for the Fig. 3 to indicate the stretched exponential decay Eq. (3) with $\beta \simeq 3/4$ (the straight line). \\
  
  {\bf B}. Anisotropic Kepler system. \\
  
\begin{figure} \vspace{-2cm}\centering
\epsfig{width=.45\textwidth,file=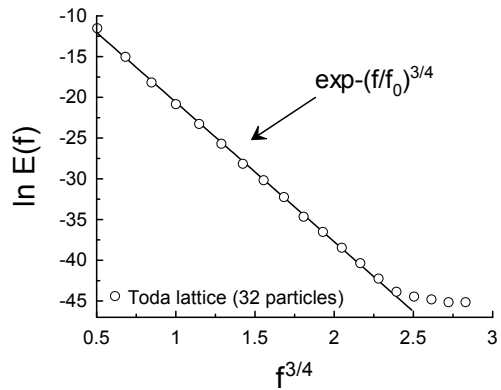} \vspace{-4cm}
\caption{A decaying part of the power spectrum of the $q_1$ coordinate for the classic Toda lattice.} 
\end{figure}
\begin{figure} \vspace{-0.5cm}\centering
\epsfig{width=.45\textwidth,file=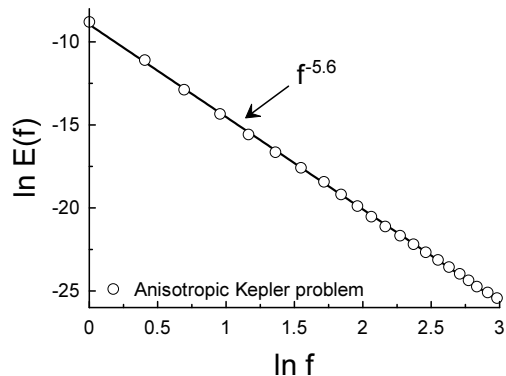} \vspace{-4.7cm}
\caption{A decaying part of the power spectrum of the $z_1$ coordinate for the anisotropic Kepler problem. } 
\end{figure}

  The anisotropic Kepler problem describes the motion of an electron in an anisotropic crystal (in the Coloumb field) \cite{rs}. Its effective mass is different along the $z$-direction and $x$-$y$ plane. In the anisotropic case the system is non-integrable and non-smooth. 
  
  Figure 4 shows spectral data taken from Ref. \cite{els2}, where the direct numerical simulations were performed for the anisotropic Kepler problem. The appropriate scales were chosen for the Fig. 4 to indicate the power-law decay Eq. (4) with $\gamma \simeq 5.6$ (the straight line). 
  
   In the Refs. \cite{li} and \cite{lig} one can find an interesting discussion on the subject and in the Refs. \cite{she}-\cite{gl2} the evidences of the relevance of the Hamiltonian dynamics to the climate.
  
\section{Large-scale climate dynamics}

 Let us start from the Arctic Oscillation. The Arctic Oscillation (AO index) is a primary annular mode of extratropical atmospheric circulation in the Northern Hemisphere \cite{tw},\cite{w}. It represents a pressure gradient between the polar and subpolar regions.

\begin{figure} \vspace{+4.5cm}\centering
\epsfig{width=.85\textwidth,file=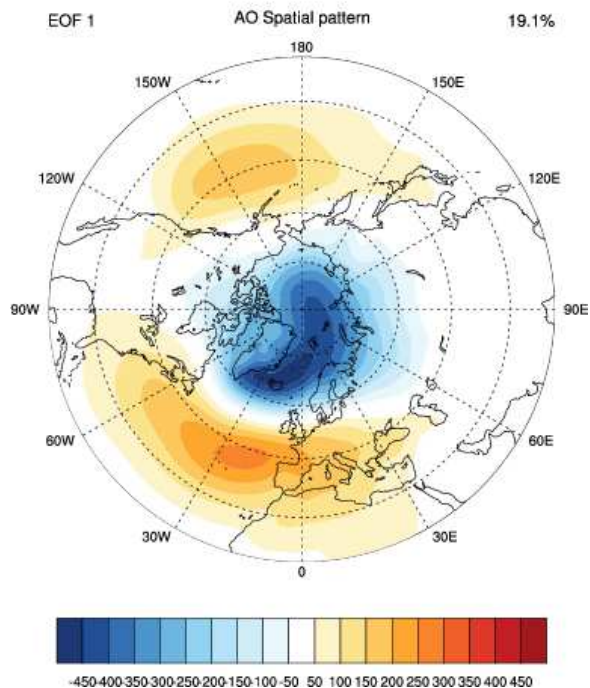} \vspace{-5cm}
\caption{Spatial pattern of the Arctic Oscillation \cite{ao1}.} 
\end{figure}
  
   At the site \cite{ao1} the {\it sea-level} pressure (SLP) based AO and AAO daily indices were calculated for a prolonged time period. The SLP is the atmospheric pressure at sea level at a given location. If a station is not located at sea level the SLP is a correction of the station pressure to corresponding sea level (taking into account the variation of pressure with height and influence of temperature).

 Figure 5 shows a spatial pattern of the Arctic Oscillation. The Empirical Orthogonal Function (EOF) is based on 1981-2010yy monthly anomalies of SLP poleward of $20^{o}$N (the 20CR V2 dataset) \cite{ao1}. The AO daily index was obtained by projecting the daily anomalies of SLP onto the EOF's. 
 
\begin{figure} \vspace{-0.5cm}\centering
\epsfig{width=.45\textwidth,file=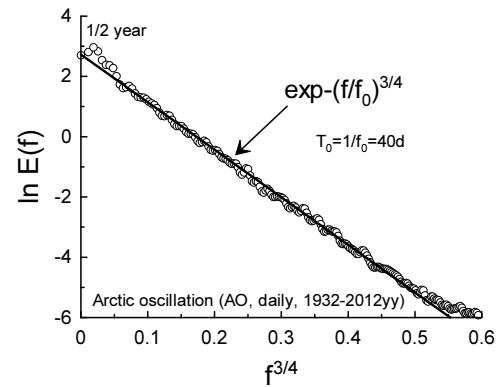} \vspace{-4.3cm}
\caption{Power spectrum of the AO daily SLP index for the period 1932-2012yy. The data for computation were taken from the site \cite{ao1}.} 
\end{figure}

\begin{figure} \vspace{-0.5cm}\centering
\epsfig{width=.45\textwidth,file=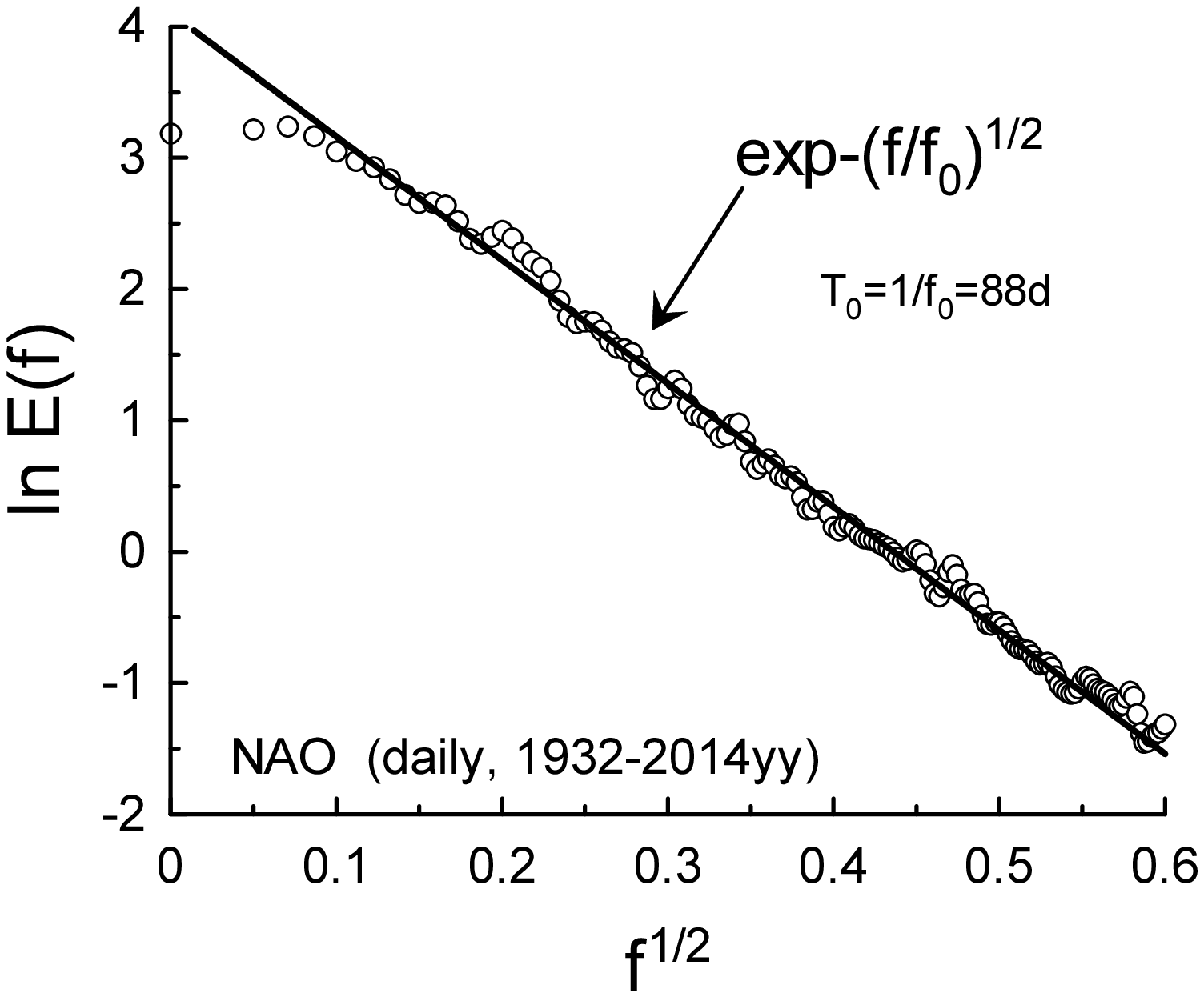} \vspace{-4.3cm}
\caption{Power spectrum of the NAO daily SLP index for the period 1932-2014yy. The data for computation were taken from the site \cite{cro}.} 
\end{figure}
\begin{figure} \vspace{-0.5cm}\centering
\epsfig{width=.45\textwidth,file=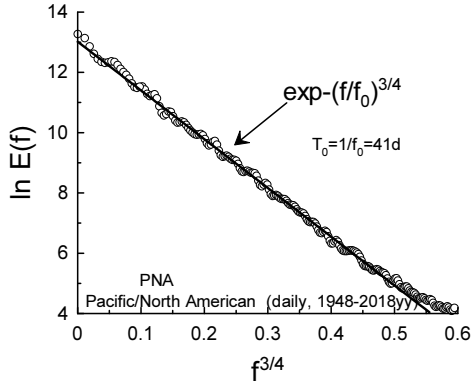} \vspace{-4.3cm}
\caption{Power spectrum of the PNA daily index for the period 1948-2018yy. The data for computation were taken from the site \cite{pna}.} 
\end{figure}
\begin{figure} \vspace{-0.5cm}\centering
\epsfig{width=.45\textwidth,file=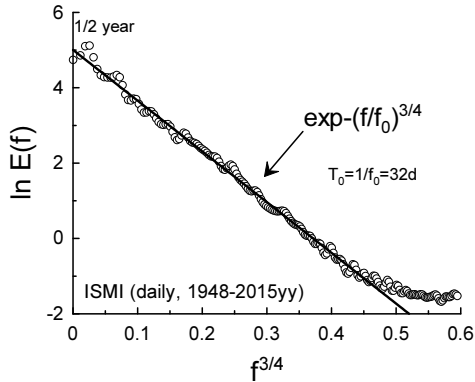} \vspace{-4.3cm}
\caption{Power spectrum of the intraseasonal daily ISM index for the period 1948-2015yy. The data for computation were taken from the site \cite{mon}.} 
\end{figure}
\begin{figure} \vspace{-0.5cm}\centering
\epsfig{width=.45\textwidth,file=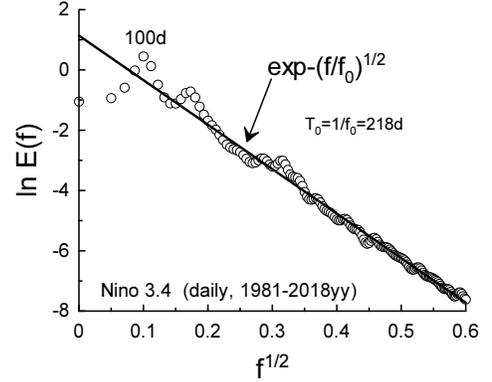} \vspace{-4.3cm}
\caption{Power spectrum of the daily Nino 3.4 index for the period 1981-2018yy. The data for computation were taken from the site \cite{cli}.} 
\end{figure}
 Figure 6 shows power spectrum computed for the daily AO index for period 1932-2012yy (the timeseries were taken from the site \cite{ao1}). The power spectrum was computed by the maximum entropy method (with an optimal resolution \cite{oh}). The straight line indicates correspondence to the Eq. (3) with $\beta = 3/4$. \\
 
   The southern counterpart of the Arctic Oscillation - the Antarctic Oscillation (AAO) exhibit analogous spectral properties (but with $\beta \simeq 1/2$). Although the corresponding data for the AAO daily index are less reliable.\\
   
     Figure 7 shows power spectrum computed for the North Atlantic Oscillation (NAO) daily SLP index for period 1932-2014yy (the timeseries were taken from the site \cite{cro}, see also Ref. \cite{cro2} for more details about the data acquisition). For a comprehensive review of the North Atlantic Oscillation properties see Ref. \cite{hur}.\\
     
     Figure 8 shows power spectrum computed for the Pacific/North American (PNA) daily index for period 1948-2018yy (the timeseries were taken from the site \cite{pna}). The computations of the PNA daily index were made for 500mb height oscillating patterns (one can find more details of the  NCEP-NCAR R1 reanalysis in Ref. \cite{ka}).\\
     
     Figure 9 shows power spectrum computed for the intraseasonal (i.e. after subtraction of the interannual variability and annual cycle) daily Indian Summer Monsoon (ISM) index for the period 1948-2015yy (the timeseries were taken from the site \cite{mon}). A comprehensive
review of the Monsoon properties can be found in Ref. \cite{wwl}. Analogous situation takes also place for the complimentary Western North Pacific Monsoon (WNPM) intraseasonal daily index as well as for their equatorial counterpart the Australian Monsoon.\\

  Figure 10 shows power spectrum computed for the intraseasonal (i.e. after subtraction of the interannual variability and annual cycle) daily Ni\~no 3.4 index characterizing the El Ni\~no/La Ni\~na phenomenon. The index is based on SST anomalies averaged across the Ni\~no 3.4 region in the east-central Equatorial Pacific (the timeseries were taken from the site \cite{cli}). The Nino 3.4 index is the most commonly used index to describe El Ni\~no/La Ni\~na events \cite{elnino}. \\
  
   One can see that for the leading large-scale climate oscillations (patterns) the decaying part of the power spectrum is a stretched exponential Eq. (3) (with the Hamiltonian values of the $\beta \simeq 3/4$ or $\beta \simeq 1/2$), that indicates the smooth predictability (it should be noted that the daily global temperature anomalies also belong to this class \cite{b4}). Only for one important large-scale climate oscillation - Madden-Julian Oscillation or MJO (the RMM and VPM daily indices \cite{wh},\cite{v}) the decaying part of the power spectrum has a power law  - Eq. (4) (with $\gamma \simeq 3$) \cite{b3}.  
  
\section{Madden-Julian Oscillation}  
  
    A sketch (by Fiona Marti, adapted from Ref. \cite{pic}) of the surface and upper-atmosphere formation of the Madden-Julian Oscillation (MJO) is shown in figure 11. The enhanced convective phase (the thunderstorm cloud) is concentrated over the Indian Ocean while the suppressed convective phase dominates atmosphere over the west-central Pacific Ocean. Traversing the planet in eastward direction this formation returns with an average period near 50 days. It is a planetary-scale coupled recurring pattern in baroclinic circulation and deep convection  \cite{v},\cite{zha}. The MJO has strong influence on extratropic weather and climate through its global teleconnections \cite{stan}-\cite{sg1}.\\
    
  Two daily indices RMM1 and RMM2 (representing two principal components \cite{wh}) are now widely used for the predicting and monitoring purposes related to the MJO dynamics. The RMM indices are based on a pair of EOFs of the near-equatorially (averaged) fields: the satellite-observed OLR (outgoing long-wave radiation), 850-hPa zonal wind and 200-hPa zonal wind. The indices were computed by projection of the data onto the EOFs (the interannual variability and annual cycle were removed from the time series \cite{wh}).
\begin{figure} \vspace{-0.5cm}\centering
\epsfig{width=.45\textwidth,file=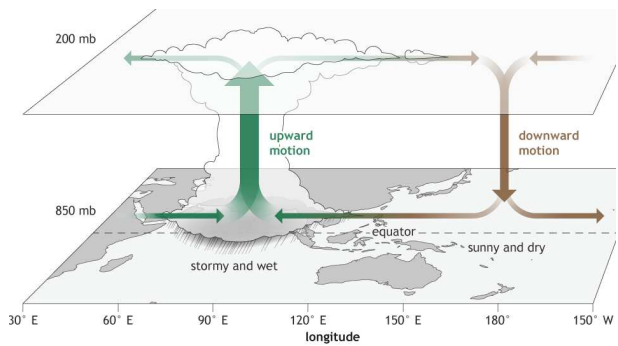} \vspace{-4.3cm}
\caption{A sketch of the surface and upper-atmosphere formation of the MJO.} 
\end{figure}
\begin{figure} \vspace{-0.5cm}\centering
\epsfig{width=.45\textwidth,file=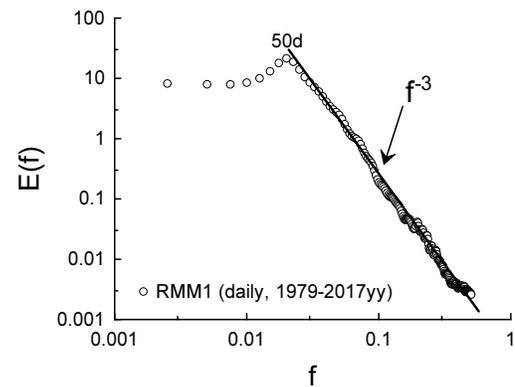} \vspace{-4.3cm}
\caption{Power spectrum of the RMM1 daily index for the period 1979-2017yy. The data for computation were taken from the site \cite{cli}.} 
\end{figure}
\begin{figure} \vspace{-0.5cm}\centering
\epsfig{width=.45\textwidth,file=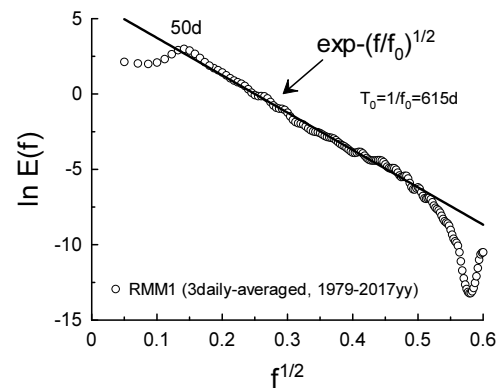} \vspace{-4.3cm}
\caption{Power spectrum of the daily RMM1 index smoothed by the three days running average.} 
\end{figure}
   Figure 12 shows power spectrum computed for the daily RMM1 index for period 1979-2017yy (the data - time series, were taken from the site \cite{cli}). The straight line indicates correspondence to the Eq. (4) with $\gamma \simeq 3$. 
   
   Therefore, unlike all the other above mentioned large-scale climate oscillations, the MJO (represented by the RMM1 index) should have rough predictability. For RMM2 index and for the complimentary daily indices - VPM1 and VPM2 the situation is similar (see the Ref. \cite{b3} for more details).
   
   But the situation can be easily improved in this case. The power law decay observed in the Fig. 12 and the corresponding rough type of the predictability can be considered as a result of a random noise produced by the high frequency waves. Such waves are common for the tropics (see, for instance, Ref. \cite{tk} and references therein). The average period of these waves is less than three days \cite{tk}. Therefore, using a simple (smoothing) filter - three days running average, we can remove the random influence of these waves on the daily time series RMM1, RMM2, VPM1 and VPM2 (cf. the Ref. \cite{b3}). Figure 13 shows power spectrum computed for the smoothed (filtered) daily RMM1 index. The straight line indicates correspondence to the Eq. (3) with $\beta = 1/2$ (one of the Hamiltonian values, it is related to spontaneous breaking of time translational symmetry, time depended Hamiltonian and action as an adiabatic invariant \cite{b3}). For the other MJO indices - RMM2, VPM1 and VPM2 the situation is similar (cf. the Ref. \cite{b3}). \\

\section{Appendix A: Chaotic thermal convection}

 Chaotic regimes in the Rayleigh-B\'{e}nard (thermal) convection were extensively studied in the experiments and numerical simulations (see, for instance, Refs. \cite{loren}, \cite{kadan}, \cite{paul},\cite{bchaos} and references therein). \\
\begin{figure} \vspace{-1.8cm}\centering
\epsfig{width=.45\textwidth,file=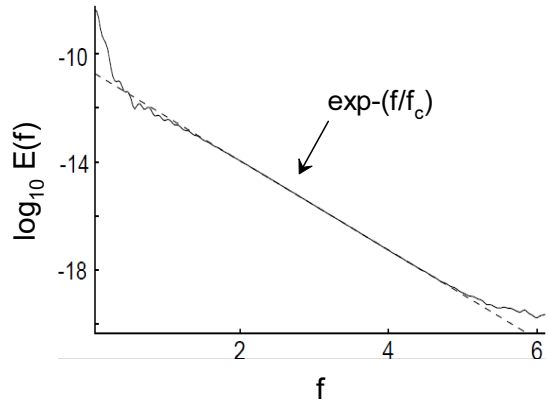} \vspace{-4cm}
\caption{Power spectrum of the dimensionless heat transport fluctuations at the onset of chaos.} 
\end{figure}
\begin{figure} \vspace{-0.5cm}\centering
\epsfig{width=.45\textwidth,file=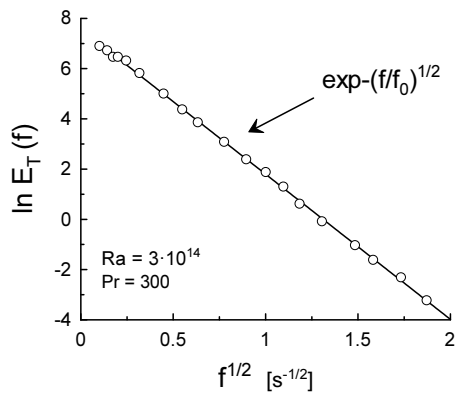} \vspace{-4.7cm}
\caption{Power spectrum of the temperature measured in the center of the domain in the strong thermal convection.} 
\end{figure}
\begin{figure} \vspace{-1cm}\centering
\epsfig{width=.45\textwidth,file=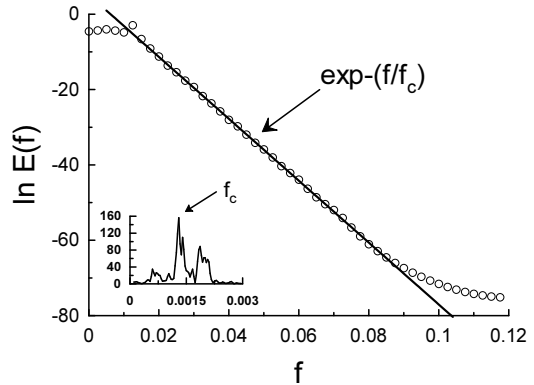} \vspace{-4.8cm}
\caption{Power spectrum of the $x$ component for the Hadley circulation model.} 
\end{figure}

   Let us start from the onset of chaos. In the paper Ref. \cite{paul} results of 3D numerical simulations in a cylindrical domain at low Rayleigh and Prandtl numbers were reported. The authors of the Ref. \cite{paul} computed power spectrum of the fluctuations of the dimensionless heat transport (across the layer) generated by the 3D Boussinesq equations. They mentioned an exponential spectral decay at the high-frequencies (see Fig. 14 adapted from the Ref. \cite{paul}) and a power-law spectral decay in a low frequencies range. The exponential spectral decay at the high frequencies the authors relate to time continuous deterministic-chaotic dynamics, whereas the power-law spectral decay at the low frequencies they relate to quasi-discontinuous dynamics.\\
   
   Unlike the previously discussed numerical simulation performed near the onset of chaos (low Rayleigh number) the laboratory experiment described in the Ref. \cite{as} was performed for very strong (turbulent) convection with the high Rayleigh and Prandtl numbers (also in a cylindrical domain). Figure 15 shows power spectrum of the temperature measured in the center of the domain. The spectral data were taken from the Ref. \cite{as}. The straight line indicates the stretched exponential decay Eq. (3) with $\beta =1/2$ (see also the Ref. \cite{kadan}).

\section{Appendix B: Low order Hadley circulation model }

  In the paper Ref. \cite{lorenz-84} Lorenz suggested a low order model of long-term atmospheric circulation. Similarly to the Lorenz system Eq. (1) this model was obtained using a Galerkin projection
$$
\begin{aligned}
\dot{x}= -ax - y^2 - z^2 + aF,~~~~~~~~~~~~~~~~~~ \\
\dot{y}= -y + xy - bxz + G,  ~~~~~~~~~~~~~(B1)  \\
\dot{z}= -z + bxy + xz       ~~~~~~~~~~~~~~~~~~~~~~~~~~~
\end{aligned}
$$

  In the model the variables $x,~y,~z$ stand for the strength of globally averaged symmetric westerly wind current, for the strength of cosine and sine phases of the travelling waves transporting heat poleward, correspondingly. The forcing terms $F$ and $G$ account for the heating contrast: symmetric cross-latitude and asymmetric between oceans and continents, correspondingly.  \\
\begin{figure} \vspace{-1.3cm}\centering
\epsfig{width=.45\textwidth,file=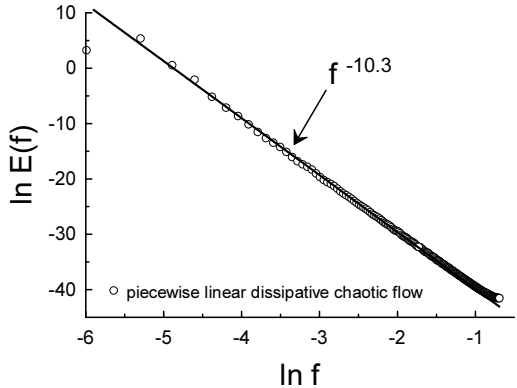} \vspace{-4.3cm}
\caption{Power spectrum of the $x$ component for the piecewise linear dissipative chaotic flow.} 
\end{figure}
\begin{figure} \vspace{-0.5cm}\centering
\epsfig{width=.45\textwidth,file=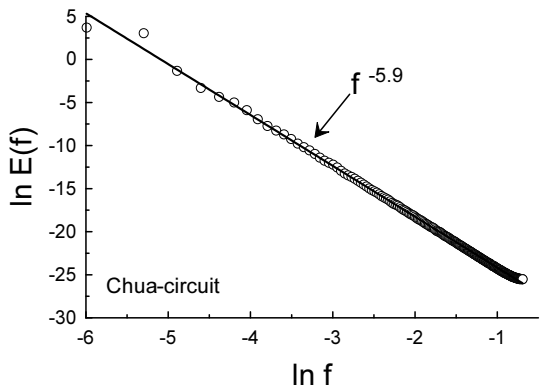} \vspace{-4.3cm}
\caption{Power spectrum of the $x$ component for the Chua system.} 
\end{figure}
  Figure 16 shows power spectrum of the $x(t)$ component for the Hadley circulation model \cite{spot}. The data for the computation were taken from the site \cite{gen}. The power spectrum was computed by the maximum entropy method (with an optimal resolution \cite{oh}). The insert shows a low frequency range of the power spectrum (computed by the fast Fourier transform method). One can see that the leading frequency $f_c$ appears also in the exponential decay. This fact indicates a strong correlation (tuning) between the fuzzy quasi-periodic coherent structures at the low frequencies and the chaotic dynamics at the high frequencies. 

\section{Appendix C: Non-smooth dynamical systems }

  The above discussed Lorenz models are represented by the smooth equations. Let us consider two examples of dynamical systems with non-smooth equations.
  
  \subsection{1C. Simplest piecewise linear dissipative chaotic flow}

In the Ref. \cite{ls} the simplest piecewise linear dissipative chaotic flow has been studied 
using system of equations:
$$
\begin{aligned}
~~~~~~~~~~~~~~~~\dot{x}= y,~~~~~~~~~~~~~~~~~~~~~~~~~~~~~~~~~~~ \\
~~~~~~~~~~~~~~~~\dot{y}= z,  ~~~~~~~~~~~~~~~~~~~~~~~~~~~~~(C1)  \\
~~~~~~~~~~~~~~~~\dot{z}= -Az -y-|x| + 1      ~~~~~~~~~~~~~~~
\end{aligned}
$$\\

  Figure 17 shows power spectrum of the $x(t)$ component for this system at $A=0.6$. The data for the computation were taken from the site \cite{gen}. One can see that unlike the smooth systems this non-smooth system exhibits a power-law spectral decay.\\

   \subsection{2C. Chua system}

   Another well known example of chaotic dynamical system with non-smooth equations is the Chua system \cite{spot}:
$$
\begin{aligned}
~~~~~~\dot{x}= a(y-h(x)),~~~~~~~~~~~~~~~~~~~~~~~~~~~~~~~~~ \\
~~~~~~~~~~~~\dot{y}= x-y+z,  ~~~~~~~~~~~~~~~~~~~~~~~~~~~~~~(C2)  \\
~~~~~~\dot{z}= -by      ~~~~~~~~~~~~~~~~~~~~~~~~~~~~~~~~~~~~~~~~~~~~~
\end{aligned}
$$
where
$$
h(x) = m_1x+\frac{1}{2}(m_0-m_1) (|x+1|-|x-1|)~~~(C3)
$$\\

  Figure 18 shows power spectrum of the $x(t)$ component for this system at $a=9.35$, $b=14.79$, $m_0 = -1/7$, $m_1=2/7$. The data for the computation were taken from the site \cite{gen}. And again one can see that unlike the smooth systems this non-smooth system exhibits a power-law spectral decay.\\

 The above mentioned correlation (tuning) between the fuzzy quasi-periodic coherent structures at the low frequencies and the chaotic dynamics at the high frequencies cannot occur in the non-smooth dynamical systems due to the scale-invariant nature of the power-law spectral decay.
  
\section{Appendix D: Systems with infinite number of degrees of freedom  }

 \subsection{1D. Soliton chaos}
\begin{figure} \vspace{-1.3cm}\centering
\epsfig{width=.45\textwidth,file=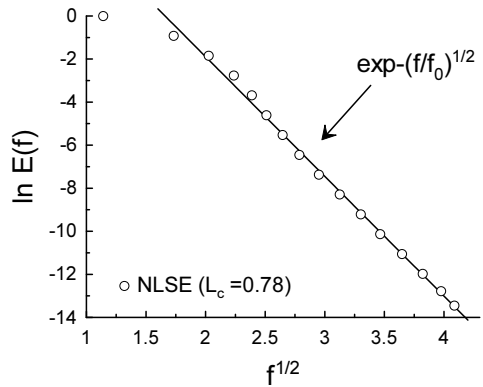} \vspace{-4.3cm}
\caption{The transverse spectrum at a slightly supercritical $L_c = 0.78$ (at $\xi=100$).} 
\end{figure}
\begin{figure} \vspace{-0.5cm}\centering
\epsfig{width=.45\textwidth,file=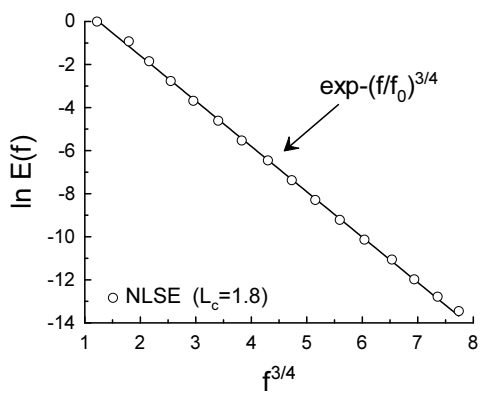} \vspace{-4.3cm}
\caption{As in Fig. 19 but for $L_c = 1.8$ (a developed soliton chaos).} 
\end{figure}
  The nonlinear Schr\"{o}dinger equation for the envelope of a physical field 
$$
{\rm i} \frac{\partial \psi}{\partial \xi} = - \frac{1}{2}\frac{\partial^2 \psi}{\partial \tau^2} - |\psi|^2 \psi,
\eqno{(D1)}
$$     
represents an integrable Hamiltonian dynamical system  \cite{ahm} with smooth predictability. The integrability of the equation Eq. (D1) results also in the existence of breathers and solitons located in the chaotic sea of the low amplitude radiation waves. The solitons are moving in all directions and their collisions produce high amplitude fluctuations of the chaotic field. 

  The authors of the Ref. \cite{ahm} reported results of a numerical simulations with the Eq. (D1) under periodic conditions for the (transverse) variable $\tau$ and random initial conditions on the variable $\xi$. The generation of solitons has been activated when the correlation length in the initial conditions $L_c$ becomes larger then a critical value. Figure 19 shows the transverse spectrum at a slightly supercritical $L_c = 0.78$ (at $\xi=100$).  The straight line in the Fig. 19 indicates the stretched exponential decay Eq. (3) with $\beta \simeq 1/2$ for the high frequencies. 
  
    Figure 20 shows the transverse spectrum at $L_c = 1.8$ (and the same $\xi=100$). At this initial value of the correlation length the soliton chaos is already a fully developed one. The straight line in the Fig. 20 indicates the stretched exponential decay Eq. (3) with $\beta \simeq 3/4$ for the high frequencies. 
    
\subsection{2D. Isotropic turbulence}
\begin{figure} \vspace{-1.8cm}\centering
\epsfig{width=.45\textwidth,file=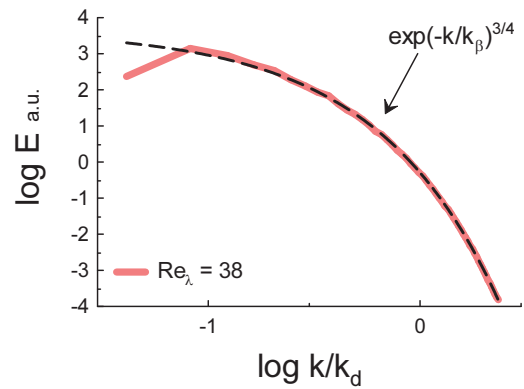} \vspace{-3.7cm}
\caption{3D spatial power spectrum at the onset of steady isotropic turbulence.} 
\end{figure}

  The Taylor hypothesis relates the spatial and temporal characteristics of turbulence. Namely, if the mean flow speed is much larger than the turbulence fluctuations intensity, then the temporal fluctuations at a fixed spatial point can be considered as the result of spatial fluctuations convected past the point by the mean velocity \cite{my}. In particular, this hypothesis allows 
to consider the spatial  (wavenumber) spectrum as an analogy of the temporal (frequency) spectrum, at least in certain ranges of the wavenumbers-frequencies. In the abstract isotropic homogeneous turbulence there is no mean velocity. Therefore the Taylor hypothesis cannot be applied to such  type of turbulence. However, the real direct numerical simulations of the isotropic turbulence are performed in the finite space domains. Therefore there always are large scale coherent structures in the simulated velocity field satisfying the condition of isotropy in a statistical sense only.    These large scale coherent structures can effectively provide applicability of the Taylor hypothesis for the spatial scales considerably lesser than the domain's scale.\\

  Figure 21 shows in the log-log scales ($\log ~k \equiv \log_{10} ~k)$ 3D spatial power spectrum at the onset of statistically steady isotropic turbulence described by the Navier-Stokes equations   
$$   
\frac{\partial \bm u(\bm x,t)}{\partial t} + (\bm u\cdot \nabla) \bm u= -\nabla p+\nu\Delta \bm u   + {\bf f}  \eqno{(D2)}
$$
$$
\nabla\cdot \bm u(\bm x,t)=0  \eqno{(D3)}.
$$    
at the Taylor microscale Reynolds number $R_{\lambda} = 38$ ($k_d$ is the Kolmogorov's or viscous scale, see the Ref. \cite{gfn} for description of the direct numerical simulations and the spectral data).  The dashed curve indicates the stretched exponential spectral decay with the wavenumber $k$
$$
E(k) \propto \exp-(k/k_{\beta})^{\beta}  \eqno{(D4)}
$$
in the log-log scales (cf. the frequency spectral decay Eq. (3)). The exponent $\beta \simeq 3/4$ in this case.
\begin{figure} \vspace{-1.6cm}\centering
\epsfig{width=.45\textwidth,file=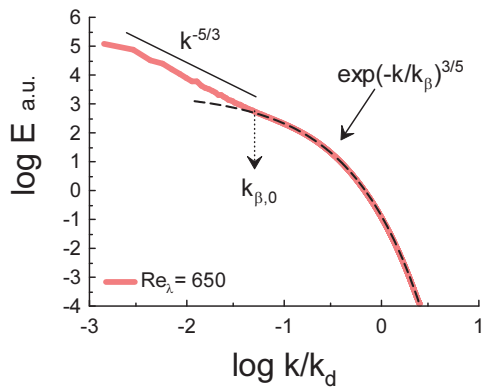} \vspace{-3.7cm}
\caption{As in Fig. 21 but for fully developed isotropic turbulence.} 
\end{figure}
\begin{figure} \vspace{-1.5cm}\centering
\epsfig{width=.45\textwidth,file=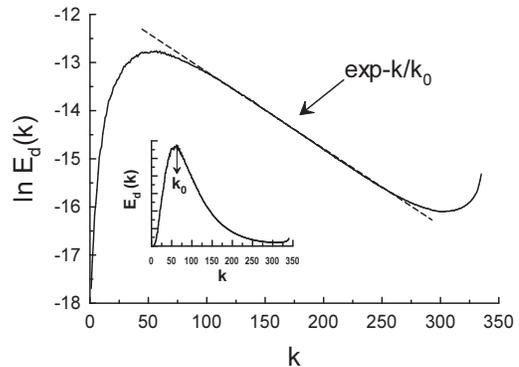} \vspace{-3.8cm}
\caption{Power spectrum $E_d(k)$ (in the semi-log scales) at $R_{\lambda} \simeq 158$).} 
\end{figure}
\begin{figure} \vspace{-0.5cm}\centering
\epsfig{width=.42\textwidth,file=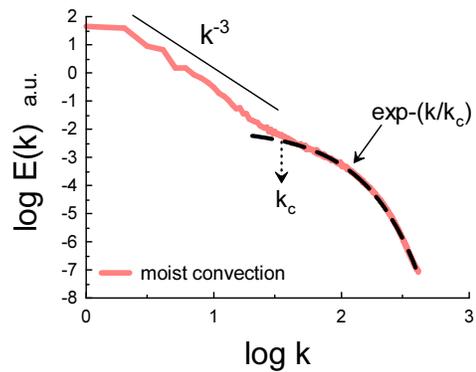} \vspace{-4.2cm}
\caption{Total power spectrum of the full-state background flow with the baroclinic waves and the moist convection.} 
\end{figure}
   Figure 22 shows 3D spatial power spectrum of the steady isotropic turbulence described by the Navier-Stokes equations at the $R_{\lambda} = 650$ (see the Ref. \cite{isy} for description of the direct numerical simulations and the spectral data). At this $R_{\lambda}$ one can see two different spectral ranges: one with the Kolmogorov power-law spectrum $k^{-5/3}$ and another with the stretched exponential decay Eq. (D4)). 
   The value $k_{\beta , 0}$ separates between the the smooth and rough ranges: $\ln k_d/ k_{\beta , 0} \simeq 3$ (see the Ref. \cite{bchaos}). In this case realization of smooth or rough predictability can depend on the spatial properties of the errors (cf. the Ref. \cite{bm}).\\

   An interesting direct numerical simulation of the steady isotropic turbulence at a small (but turbulent) $R_{\lambda} \simeq 158$ was described in the recent Ref. \cite{bh}.  
   
  A copy of the velocity field $\bm{u}_1$ was perturbed using a slight perturbation of the
forcing function ${\bf f}$ at one particular timestep. A new field ${\bf u}_2$ was created by this perturbation. The forcing was applied at the low wavenumbers only.

  The error field $\delta \bm{u} = \bm{u}_1 - \bm{u}_2$ was then computed as well as its power spectrum, $E_d(k,t)$
$$
E_d(k,t) = \frac{1}{2} \int_{|\bm{k}|=k} d \bm{k} 
|\bm{\hat{u}}_1 ( \bm{k},t) -  \bm{\hat{u}}_2 ( \bm{k},t)|^2 \  \eqno{(D5)}
$$

 Figure 23 shows the power spectrum $E_d(k)$ (in the semi-log scales) after a steady state was reached. The dashed straight line indicates an exponential spectral decay. As one can see from the insert the peak of the $E_d(k)$ spectrum is reached at the same value of $k\simeq k_0$ that appears in the exponential spectral decay. This fact indicates a tuning between the large-scale coherent structures at the comparatively low wavenumbers and the chaotic dynamics at the high wavenumbers. 
 
\subsection{3D. Baroclinic waves chaos with and without moist convection}

  In recent Ref. \cite{sz} the idealized baroclinic waves' instabilities with and without moist convection were studied using the Advanced Research version of the Weather Research and Forecast Model \cite{ska},\cite{wz}. Figure 24 shows a total power spectrum of the full-state background flow for the baroclinic waves with moist convection. One can see two different spectral ranges: one with the power-law spectrum $k^{-3}$ and another with an exponential decay. A numerical support of rough predictability has been provided in the Ref. \cite{sz} for this flow, that seems to be in accordance with results of the recent direct numerical simulations reported in the Ref. \cite{leu} (which relates the $k^{-3}$ spectral decay to the rough predictability, see also site https://presentations.copernicus.org/EGU2018-371\_presentation.pdf).  
  
    Figure 25 shows a total power spectrum of the full-state background flow for the baroclinic waves without moist convection (the 'dry' case in the Ref. \cite{sz}). The dashed straight line indicates a stretched exponential spectrum Eq. (D4) with unusually small value of $\beta \simeq 1/7$. It should be noted that the entire range of the wavenumbers shown in the Fig. 25 is the same as the entire range of the wavenumbers shown in the Fig. 24 (i.e. about two and half decades).  A numerical support of smooth predictability has been provided in the Ref. \cite{sz} for this flow, as it can be expected if the Taylor hypothesis can be applied in this case.
\begin{figure} \vspace{-1.1cm}\centering
\epsfig{width=.42\textwidth,file=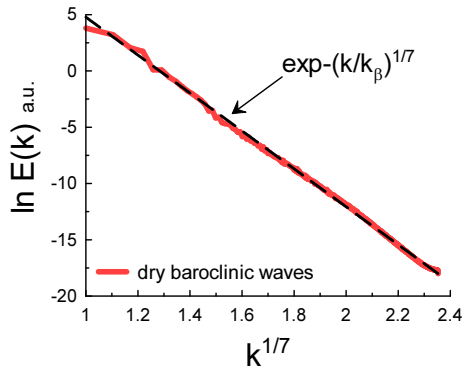} \vspace{-3.6cm}
\caption{As in Fig. 24 but for the dry baroclinic waves.} 
\end{figure}

\section{Appendix E: Perturbations dynamics in ensemble weather forecasting}
\begin{figure} \vspace{-0.9cm}\centering
\epsfig{width=.42\textwidth,file=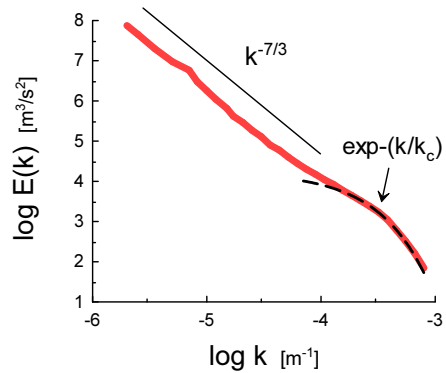} \vspace{-3.8cm}
\caption{Kinetic energy spectrum for the control run.} 
\end{figure}
\begin{figure} \vspace{0cm}\centering
\epsfig{width=.42\textwidth,file=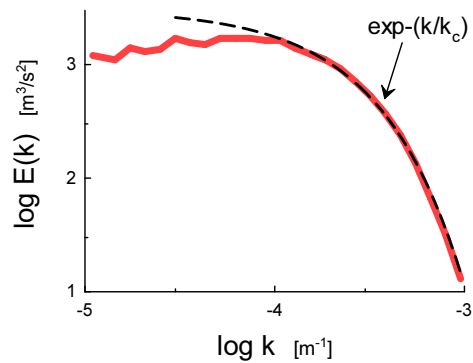} \vspace{-4cm}
\caption{Kinetic energy spectrum of the
difference between the control run and perturbed runs at 12 hours of perturbation lead time (the spectrum is averaged over all four perturbation runs).} 
\end{figure}
\begin{figure} \vspace{-1.3cm}\centering
\epsfig{width=.42\textwidth,file=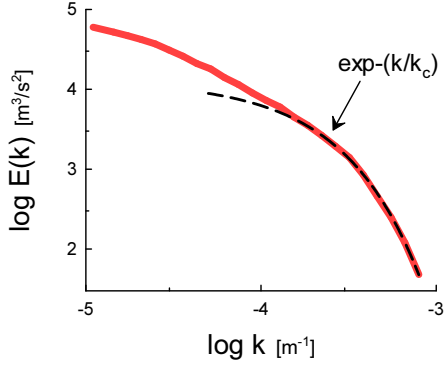} \vspace{-3.8cm}
\caption{As in the Fig. 27 but at 60 hours of perturbation lead time.} 
\end{figure}

  The authors of a recent paper Ref. \cite{sg2} simulated upscale error growth in a weather event using the Consortium for Small-Scale Modeling (COSMO) model (see also Ref. \cite{bal}).  At this event several days (19-23 July 2007) of cold front passage and intense convective activity spanning over central Europe were caused by about stationary low pressure system located at Great Britain.

  The simulation was started at 0000 UTC 19 July with real-atmospheric data.  The boundary and initial conditions were derived from the European Centre for Medium- Range Weather Forecasts (ECMWF) analysis and deterministic forecast. First (control) run of the simulation was performed without any perturbation for 96 hours period. Four additional runs with a single-time tiny bit perturbation, imposed at 15, 21, 27, and 33 hours after forecast beginning, were also performed in order to generate different meteorological realizations. A Gaussian uncorrelated small-amplitude and grid-scale noise was added to the temperature field at these perturbations. \\

  Figure 26 shows in the log-log scales kinetic energy spectrum for the control run (the spectral data were taken from the Fig. 4 of the Ref. \cite{sg2}). The straight line indicates the power-law spectral decay $k^{-7/3}$ typical for the quasi two-dimensional case \cite{bkt}. The dashed curve indicates an exponential spectral decay (cf. Figs. 24).
  
   Figure 27 shows in the log-log scales kinetic energy spectrum of the
difference between the control run and perturbed runs at 12 hours of perturbation lead time (the spectrum was averaged over all four perturbation runs). The dashed curve indicates the exponential spectral decay. One can see that the exponential covers entire decaying part of the spectrum in this case.

  Figure 28 shows in the log-log scales kinetic energy spectrum of the
difference between the control run and perturbed runs at 60 hours of perturbation lead time (the spectrum was averaged over all four perturbation runs). The dashed curve indicates the exponential spectral decay.

  It is emphasised in the Ref. \cite{sg2} that the observed upscale error growth and sensitivity of ensemble forecasts to large-scale vs. small-scale uncertainty can be important in limiting predictability. Considering the Section I of present paper and taking into account the Taylor hypothesis one can come to the same conclusion.\\
  
   In another recent paper Ref. \cite{dg} results of 100-member ensembles forecast for an East Coast winter storm were reported. An ensemble Kalman filter \cite{wh2} was used in order to construct the 100-member ensembles which were integrated (with
slightly altered initial conditions) for 36 hours ensemble forecast using COAMPS - Coupled Ocean–Atmosphere Mesoscale Prediction System \cite{hod}.

\begin{figure} \vspace{-1.5cm}\centering
\epsfig{width=.42\textwidth,file=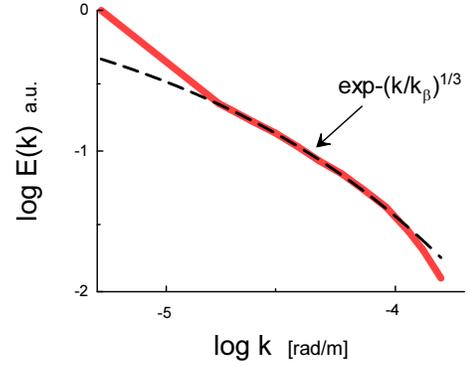} \vspace{-3.6cm}
\caption{Background kinetic energy spectrum (ensemble- and
meridional-averaged for the 25 Dec 2010 storm).} 
\end{figure}
\begin{figure} \vspace{-0.5cm}\centering
\epsfig{width=.42\textwidth,file=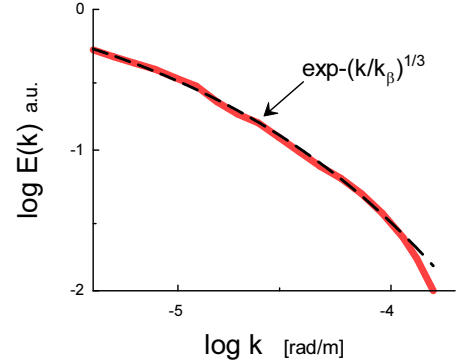} \vspace{-4.3cm}
\caption{Error kinetic energy spectrum (ensemble- and
meridional-averaged for the 25 Dec 2010 storm) at 36 hours of the lead time.} 
\end{figure}
\begin{figure} \vspace{-1.5cm}\centering
\epsfig{width=.42\textwidth,file=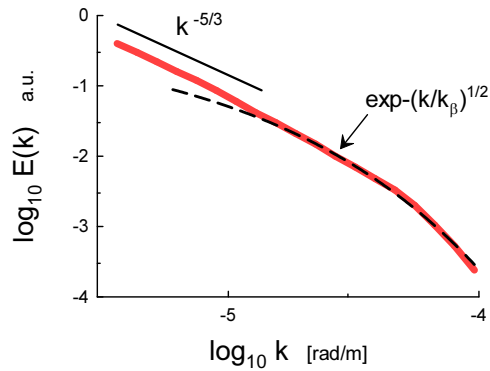} \vspace{-4cm}
\caption{Average of the horizontal kinetic energy spectra at 700-hPa at 1200 UTC 17 Dec 2008. } 
\end{figure}

\begin{figure} \vspace{-0.5cm}\centering
\epsfig{width=.42\textwidth,file=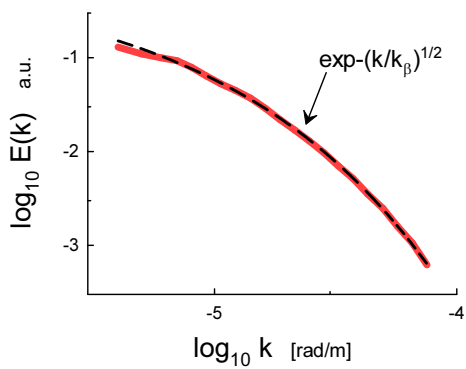} \vspace{-3.8cm}
\caption{Spectrum of the kinetic energy of the perturbations about the ensemble mean at the 36 hours of the lead time for the 17 Dec 2008 storm.} 
\end{figure}
   Figure 29 shows in the log-log scales the mean background kinetic energy spectrum (ensemble- and meridional-averaged) at 500 hPa. The simulation was started at 1200 UTC 25 Dec 2010 (cyclogenesis for the storm) with real-atmospheric data. Figure 30 shows perturbation kinetic energy spectrum (ensemble- and meridional-averaged) at the 36 hours of the lead time (the spectral data were taken from the Fig. 6b of the Ref. \cite{dg}). In this case the perturbations were initially generated (initial error).  The perturbation is the difference between the ensemble mean and one ensemble member. The dashed curve indicates the stretched exponential decay Eq. (D4) with $\beta =1/3$ (known for the buoyancy driven distributed chaos \cite{bu}). The Fig. 30 (see also Fig. 32) confirms the idea \cite{dg},\cite{drd}-\cite{map} that the error growth can be mainly result of amplification of the errors at all wavenumbers.
     
   Another winter storm in the Puget Sound region of the Pacific Northwest was studied by the same method in the Ref. \cite{drd}. Figure 31 shows average of the horizontal kinetic energy
spectra at 700-hPa at 1200 UTC 17 Dec 2008 (the spectral data were taken from the Fig. 13 of the Ref. \cite{drd}). Figure 32 shows spectrum of the kinetic energy
of the perturbations about the ensemble mean at 700 hPa at the 36 hours of the lead time (the spectral data were taken from the Fig. 14d of the Ref. \cite{drd}). The simulation was started at 0000 UTC 17 Dec 2008 with real-atmospheric data. The dashed curve indicates the stretched exponential decay Eq. (D4) with $\beta =1/2$.

\section{Acknowledgement}

I thank O. Adam, M.H.P Ambaum, B. Galperin, Y.C. Li, A. Pikovsky, S. Vannitsem, C. Varotsos and participants of the "Climate, Atmosphere and Ocean" seminar (HUJI) for stimulating comments and discussions. I also grateful to A. Berera, R.D. J. G. Ho, T. Nakano, D. Fukayama, T. Gotoh, K.P. Iyer, J.C. Sprott, Y.Q. Sun and F. Zhang for sharing their data and discussions.  I acknowledge use of the data provided by the Earth system research laboratory (NOAA), the Zenodo database (CERN), the KNMI Climate Explorer (Netherlands) and the Asia-Pacific Data-Research Center at University of Hawaii.

\end{document}